# Extremely Low Drift in Amorphous Phase Change Nanowire Materials

Mukut Mitra, Yeonwoong Jung, Daniel S. Gianola, and Ritesh Agarwal*

Department of Materials Science and Engineering, University of Pennsylvania

3231 Walnut Street, Philadelphia, PA 19104-6272.

Time-dependent drift of resistance and threshold voltage in phase change memory (PCM) devices is of concern as it leads to data loss. Electrical drift in amorphous chalcogenides has been argued to be either due to electronic or stress relaxation mechanisms. Here we show that drift in amorphized $Ge_2Sb_2Te_5$ nanowires with exposed surfaces is extremely low in comparison to thin-film devices. However, drift in stressed nanowires embedded under dielectric films is comparable to thin-films. Our results shows that drift in PCM is due to stress relaxation and will help in understanding and controlling drift in PCM devices.

* riteshag@seas.upenn.edu

Electric-field induced structural phase transformations in chalcogenides have attracted significant interest due to their potential use in non-volatile phase change memory (PCM) devices [1]. Chalcogenide materials (e.g., Ge-Sb-Te alloys) are particularly important for PCM owing to their fast and reversible crystalline to amorphous phase change properties via Joule heating to produce electrically distinct states. However, various challenges need to be addressed before PCM can become a viable alternative to flash technology. These challenges include understanding their structural, electronic, thermal and mechanical properties especially in the amorphous state, the effect of field on electrical switching, and device scalability.

Chalcogenide glasses are affected by relaxation processes that occur on a large distribution of timescales resulting in time dependent electrical [2-5], optical [6], mechanical [7], and thermal [8] behavior. For PCM device operation, the amorphous state resistance and the field strength at which it switches to a higher conducting state (threshold voltage, $V_{th}$) are fundamental parameters that determine the device reliability. Any phenomenon that affects these critical parameters leading to temporal drift is an important issue that needs to be investigated. Time dependent drift [2-4] of these parameters have important implications because they lead to changes in the measurable device parameters used for recording and reading the information in phase change materials. Particularly, the issue of drift becomes very important for designing multilevel memory devices, as drifts in resistance would lead to states being overwritten causing serious errors [9,10]. Therefore, it becomes important to understand the physical origin of phenomena that lead to drift in material properties to eventually minimize such effects.

The drift of amorphous phase resistance in PCM devices has been described by a power law (typical of activated processes in chalcogenide glasses)

$$R(t) = R(t_0)(t/t_0)^\alpha, \qquad (1)$$

where α is reported in the range of 0.03 to 0.1 depending on the device type and initial amorphous state resistance [2-4]. However, the drift in $V_{th}$ has been fitted either with a power law [2, 3], or to a logarithmic dependence [4],

$$\frac{\Delta V_{th}(t)}{V_{th}(t_0)} \equiv \frac{V_{th}(t) - V_{th}(t_0)}{V_{th}(t_0)} = \nu \ln \frac{t}{t_0}, \quad (2)$$

where υ has been reported in the range of ~0.02-0.04 [4].

The fundamental understanding and the origin of drift in phase change materials is still being debated and has been explained by a variety of effects including stress relaxations in the amorphous material [2, 4], relaxation processes which anneal the electronic defects [2,3], and the formation of valence alternation pairs (VAP) [11-13], all of which can increase the mobility gap causing an increase in material's resistivity and $V_{th}$. Within the electronic structural relaxation model, the annealing of traps occurs due to atomic motions resulting in a decrease in the inter-trap distance causing a change from modified Poole to Poole-Frenkel (P-F) mechanism of conduction. The stress relaxation model is based on the large internal hydrostatic pressure that is built in the amorphous dome due to the melt-quench process owing to the large difference between the densities of the crystalline and amorphous phase (several percent). The slow relaxation of internal stress (volume dilation), consistent with the stress relaxation data [14], is due to atomic motions within the completely embedded amorphous dome which increases the Fermi level from the valence band edge causing time-dependent increase in resistance and $V_{th}$.

One way to distinguish between these proposed mechanisms is to design experiments on un-embedded nanoscale systems in which the stress upon amorphization can relax efficiently. The stress relaxation efficiency can also be engineered by changing the surface to volume ratio or by embedding the nanostructured device without altering the material's electronic properties.

If the stress relaxation mechanism dominates the drift dynamics, then any change in the system's ability for efficient stress relaxation should significantly influence its drift characteristics.

Phase change nanowires (NWs) are an important class of materials in this regard as they represent the minimum dimension that can be reliably connected into devices, their sizes can be controlled down to 10 nm to tune the surface to volume ratio, and unlike thin-films, NW devices can be configured with their surfaces exposed or completely embedded [15-17]. Experiments on NWs have shown efficient electrical switching due to material confinement and current localization, scalability [15], multi-bit operation [16], and evidence of strong heterogeneous nucleation from its surfaces [17]. Here, we utilize the unique geometry of phase change NWs to understand the mechanism of resistance and $V_{th}$ drift in PCM. It is shown that NW devices have extremely small values of drift coefficients in comparison to thin-films. By systematically varying the stress relaxation parameters such as the surface to volume ratio of the NWs and comparing un-embedded and embedded devices, it is demonstrated that the release of the built-in stress upon amorphization is primarily responsible for drift in phase change materials.

Phase change $Ge_2Sb_2Te_5$ NWs were synthesized using the bottom-up approach of catalyst mediated vapour–liquid–solid process [15]. All NW devices were fabricated with Pt electrodes with a separation of 4 μm [15] and their resistances were measured at 0.2V (d.c.). The NW devices were amorphized by 200 ns electrical pulses to a resistance value of 1-4 MΩ, and then were allowed to relax over 5 decades of time during which the resistance was measured at different intervals. For $V_{th}$ drift measurements, d.c. *I-V* sweeps were measured beyond the $V_{th}$ and the devices were amorphized back to the original resistance for time-dependent measurements. The amorphized region in a NW device mostly occurs near the center of the device and sometimes next to the contacts, but never totally embedded under the contacts.

The time evolution of amorphous state resistance (normalized at t=1s, first measurement) of a 100 nm thick NW device from the initial value of 2.1 MΩ (Fig. 1a) clearly shows that the resistance drift is very small with the resistance increasing to just 2.3 MΩ over five decades in time. The data are fit to a power law (Eqn. 1) and the power exponent, $\alpha$ is found to be 0.005, which is very small in comparison to thin-film devices where α typically ranges from 0.03-0.1 [4]. The corresponding data for drift in $V_{th}$ (normalized at 2s, first measurement) can be fit to Eqn. 2 (Fig. 1b), which also reveals a lower value of drift exponent (υ = 0.009) in comparison to typical values ranging from 0.01-0.04 in thin-film devices [4]. These experiments demonstrate that drift coefficients are lower in NW devices in comparison to thin-film devices.

In order to study if there is any correlation of drift coefficients with NW size, we measured the drift of device resistance on three different NW thicknesses, 140, 90 and 45 nm, all amorphized to resistance values that were at least 100 times more than their crystalline state resistance. The NW devices show a systematic size-dependent drift of amorphous state resistance (Fig. 2); the drift coefficients increase with increasing NW thickness. The thinnest NW (45 nm) with the highest surface to volume ratio shows the lowest value of drift (α=0.002), while the thickest NW device (140 nm) with the lowest surface to volume ratio shows a much higher drift coefficient (α=0.009), but still much smaller than the reported values for thin-film devices The drift coefficients of $V_{th}$ for different NW thicknesses did not reveal any clear size-dependence, mostly because the $V_{th}$ changes typically from ~1.5 V (t=2s) to ~1.7 V (t=$10^5$s) for all the measured devices, a small increase to reliably extract their size-dependence.

The above described data suggests that the unique geometry of NW devices with exposed surfaces may be responsible for the observed low but size-dependent drift coefficients. In order to explore the effect of exposed surfaces on the drift behavior of NW devices, we performed the drift experiments on the same NW devices by depositing thick (~300 nm) dielectric films such as $SiO_2$ or $Si_3N_4$ on them. The devices were imaged with SEM to ensure that they were completely embedded. The same 100 nm NW device as discussed in Fig. 1, but embedded under $SiO_2$ film was amorphized again to a resistance value of 2.1 MΩ and its drift behavior was measured. The NW device now showed a much higher resistance ($\alpha$=0.086, compared to 0.005 for un-embedded), and $V_{th}$ drift coefficients ($\upsilon$ = 0.031, compared to 0.009 for un-embedded), which are similar to values reported for embedded thin-film PCM devices (Fig. 1). Similar values of increase of drift coefficients for NWs embedded under $Si_3N_4$ film in comparison to un-embedded devices were obtained (data not shown). These data suggest that the drift characteristics of the devices can be engineered by both altering the materials surface to volume ratio and exposing/embedding the surfaces.

Our measurements on NW devices clearly demonstrate the difference in drift behavior as a function of surface to volume ratio and exposed surfaces. These observations suggest that the efficient relaxation of the built-in stress upon amorphization is primarily responsible for drift in PCM devices in comparison to annealing of electronic defects or VAP generation mechanisms. In a conventional thin-film PCM device, a polycrystalline film is sandwiched between two electrodes; upon amorphization, an embedded amorphous dome results, that is encapsulated from all sides and is compressed under large stresses from the surrounding materials. In the course of time, the metastable amorphous region relaxes with a large distribution of timescales, which has been mapped with stress relaxation experiments [14], resulting in a time-dependent increase of resistance and $V_{th}$ due to volume dilation. These relaxations have been extensively studied in glasses and have been attributed to atomic structural relaxations [18,19], where unsaturated, stretched and distorted bonds characteristic of the metastable state relax to more stable states leading to time-dependent mechanical, electrical and optical properties.

Unlike thin-film devices, NW devices do not have a completely embedded amorphized dome, which will enable them to efficiently relax their stress rapidly from the large available surface (Fig 3), leading to very different drift behavior as observed. Typically, phase change NWs have a ~1-2 nm coaxial surface oxide shell and the amorphized region will terminate at this interface. This extremely thin oxide shell can easily expand when the NW device is locally amorphized thereby releasing the stress that builds up due to the material expansion. Smaller diameter NWs have large surface to volume ratio and can relax their stress more easily. It is known that NWs can be epitaxially grown on highly mismatched substrates and their heterostructures can be created that can withstand large strains due to the presence of free surfaces allowing for lateral strain relaxation [20,21]. Recent theoretical efforts on mechanical properties of NWs have also revealed that at the nanoscale, strain relaxation becomes more efficient with decreasing surface to volume ratio [22]. For the case of NW devices which are completely buried under a uniform dielectric layer, the situation becomes similar to convention thin-film devices and the amorphous volume cannot relax the built-in pressure and hence relaxes slowly on multiple timescales leading to higher drift coefficients. This is in agreement with measurements of crystallization-induced stress in $Ge_2Sb_2Te_5$ thin films where the stress was reported to be much higher for capped films [23]. It is unlikely that the deposition of the dielectric film significantly alters the electronic properties of the NWs as the presence of the thin coaxial oxide layer minimizes direct interaction of the deposited material with the active region of the NW. In addition, the drift properties of the embedded NW devices do not appear to be strongly dependent on the chemical composition of the dielectric film, although the drift dependence on the film thickness and mechanical properties would require further investigation.

We have estimated the elastic stresses associated with the amorphization of the NW geometry for both the embedded and un-embedded devices using finite element analysis to provide insight on the relationship between drift and structural confinement. Plane strain 2D calculations were performed on a $Ge_2Sb_2Te_5$ NW cross-section (100 nm x 100 nm) on a $SiO_2$ substrate, where the bottom surface of the NW was constrained while the other surfaces were modeled as free (Fig. 3b). A volume change of $\Delta V/V = tr(\varepsilon_{ij}) = 0.05$ where $\varepsilon_{ij}$ is the strain tensor, representing the decrease in density upon amorphization, Young's modulus of 35 GPa and Poisson's ratio of 0.3 were used for the $Ge_2Sb_2Te_5$ NW [24]. The first principal stress $\sigma_1$ for the un-embedded NW (Fig. 3b), reveals that a very small but finite stress develops due to the constraint from the substrate. In contrast, large compressive stresses develop when $SiO_2$ and $Si_3N_4$ capping layers are applied (Figs. 3c, 3d) due to confinement effects upon amorphization, which are on average ~20 times more than the un-embedded case. The low and size-dependent drift coefficients are consistent with the notion of the high surface to volume ratio of the NW serving to effectively accommodate the stresses, although a detailed understanding of the atomic relaxation mechanisms in $Ge_2Sb_2Te_5$ NWs requires future investigation.

One of the arguments in favor of the defect annealing model has been the observation of modified Poole mechanism of conduction (interacting traps due to high density) in some amorphized thin-film devices below $V_{th}$, which after relaxation switches to P-F conduction mechanism (isolated traps after their density decreases upon annealing/relaxation) [25]. However, all our NW devices showed P-F mechanism of conduction at all times after amorhization (Fig. 4), as demonstrated by *ln(I) α $V^{1/2}$* dependence, typical of P-F mechanism. No observable change in the I-V characteristics was observed for embedded NW devices in comparison to un-embedded devices. Thus amorphous phase PCM NW devices have low donor/trap densities that do not interact electronically, which further suggests that drift behavior of PCM parameters is not due to defect annealing, but instead to atomic structural relaxations leading to release of built-in confinement stresses.

In conclusion, we have shown that stress relaxation mechanisms are primarily responsible for the observed drift behavior in PCM devices by using un-embedded NWs as a model system. The NW devices have extremely low values of drift coefficients due to efficient stress relaxation from the exposed free surfaces in comparison to thin-film devices. These results help understand the effect of stress relaxation on electrical properties of phase change chalcogenides and will be useful for the development of new device geometries that can efficiently relax stress upon amorphization to minimize drift for practical memory applications.

Acknowledgements: This work was supported by grants from ONR (Grant # N000140910116), Materials Structures and Devices Center at MIT, NSF (DMR-0706381), and Penn-MRSEC (DMR05-20020).

**References**:

[1]     S. Raoux, Annu. Rev. Mat. Res. **39**, 25 (2009).

[2]     A. Pirovano *et al.,* IEEE Trans. Electron Devices **51**, 714 (2004).

[3]     D. Ielmini, L. Lacaita, and D. Mantegazza, IEEE Trans. Electron Devices **54**, 308 (2007).

[4]     I. V. Karpov *et al.,* J. App Phys. **102**, 124503 (2007).

[5]     R. T. Johnson, and R. K. Quinn, J. Non-Cryst. Sol. **28**, 273 (1978).

[6]     V. G. Karpov and M. Grimsditch, Phys. Rev. B **48**, 6941 (1993).

[7]     O. B. Tsiok *et al*., Phys. Rev. Lett. **80**,  999 (1998).

[8]     M. T. Loponen *et al.,* Phys. Rev. B **25**, 1161 (1982).

[9]     S. Braga, A. Cabrini and G. Torelli, App. Phys. Lett. **94**, 092112 (2009).

[10]    D. H. Kang *et al.,* IEEE Symp. VLSI Tech. Dig. Tech. Pap. 98 (2008).

[11]    R. A. Street and N. F. Mott, Phys. Rev. Lett. **35**, 1293 (1975).

[12]    M. Kastner, D. Adler, and H. Fritzsche, Phys. Rev. Lett. **37**, 1504 (1976).

[13]    M. Kastner and H. Fritzsche, Philos. Mag. B **37**, 199 (1978).

[14]    J. Kalb, F. Spaepen, T. P. L. Pedersen, and M. Wuttig, J. Appl. Phys. **94**, 4908 (2003).

[15]    S. H. Lee, Y. Jung, R. Agarwal, Nat. Nanotechnol. **2**, 626 (2007).

[16]    Y. Jung, S. H. Lee, A. T. Jennings, and R. Agarwal, Nano Letts. **8**, 2056 (2008).

[17]    S. H. Lee, Y. Jung, R. Agarwal, Nano Letts. **8**, 3303 (2008).

[18]    P. W. Anderson, B. I. Halperin, and C. M. Varma, Philos. Mag. **25**, 1 (1972).

[19]    W. A. Phillips, J. Low Temp. Phys. **7**, 351 (1972)

[20]    P. Caroff  *et. al*., Small **4**, 878 (2008).

[21]    T. Martensson et al.,Adv. Mater. **19**, 1801 (2007).

[22]    H. S. Park and P. A. Klein, Phys. Rev. B. **75**, 085408 (2007).

[23]    Q. Guo *et al.,* Appl. Phys. Lett.  **93**, 221907 (2008).

[24]    T.P. Leervard Pederson *et al.*, Appl. Phys. Lett. **79**, 3597 (2001).

[25]    D. Ielmini and Y. Zhang, J. Appl. Phys. **102**, 054517 (2007).

**Figure Captions:**

**Fig. 1.** (a) Time-dependent normalized resistance drift behavior of a 100 nm thick un-embedded $Ge_2Sb_2Te_5$ NW device (triangles), and the same device embedded under a 300 nm thick $SiO_2$ film (squares), amorphized to a resistance of 2.1 MΩ. The solid lines are fit to Eqn. 1 and the corresponding drift coefficients are given (b) Drift of normalized $V_{th}$ for an un-embedded (triangles) and the embedded NW device (squares). Solid lines are fit to Eqn. 2. The embedded device shows higher drift coefficients in comparison to un-embedded devices.

**Fig. 2.** Size-dependent drift of normalized resistance for $Ge_2Sb_2Te_5$ NW devices. Smaller diameter devices show less drift compared to larger diameter devices. Solid lines are fit to Eqn. 1 and all devices were amorphized to resistance ratios of at least 100 in comparison to their crystalline state.

**Fig. 3.** (a) SEM image of an un-embedded NW device. (b-d) finite element analysis of the elastic first principal stresses that develop in a $Ge_2Sb_2Te_5$ NW cross-section (100 nm x 100 nm) upon incurring a volume change of 5%. An un-embedded NW (b) develops small stresses due to adhesion at the substrate interface, but large compressive stresses develop in devices confined with $SiO_2$ (c) and $Si_3N_4$ (d). These stresses are likely accommodated by relaxation mechanisms in the amorphous material, which influences the drift behavior [14]. Note that stresses in the substrate and capping layers are present but have been omitted for clarity.

**Fig. 4.** Sub-threshold current-voltage (*I-V*) plots for three amorphized NWs devices (from Fig. 2) at room-temperature measured 1 s after amorphization. Each of those plots can be fitted with $ln(I) \alpha V^{1/2}$, which suggests Poole-Frenkel conduction mechanism corresponding to non-interacting electronic traps. The *I-V* characteristics did not change significantly with time.

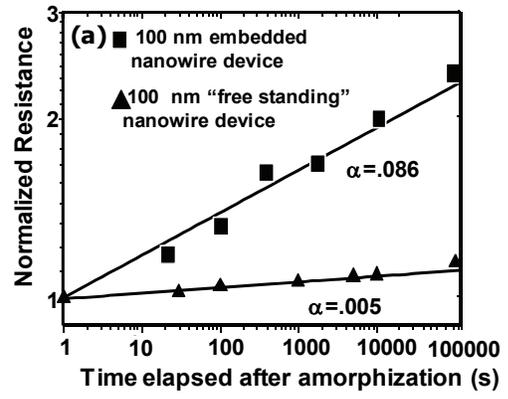

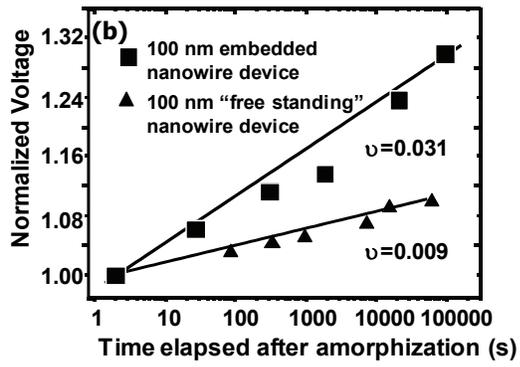

Figure 1, Mitra et al.

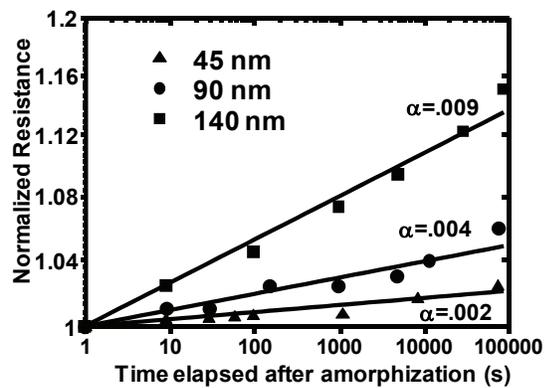

Figure 2, Mitra et al.

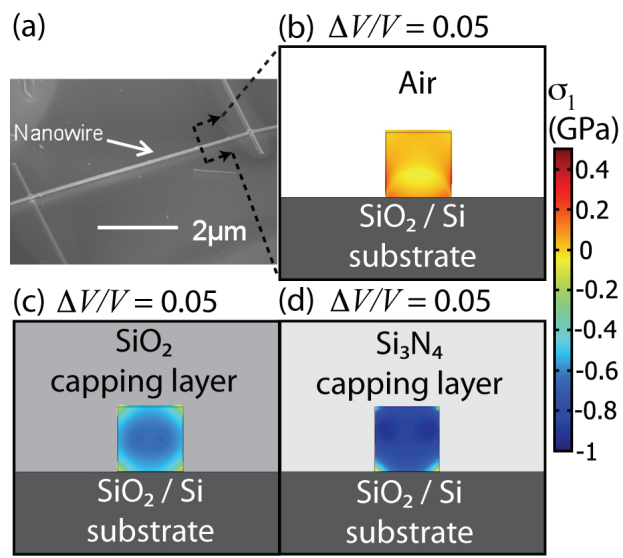

Figure 3, Mitra et al.

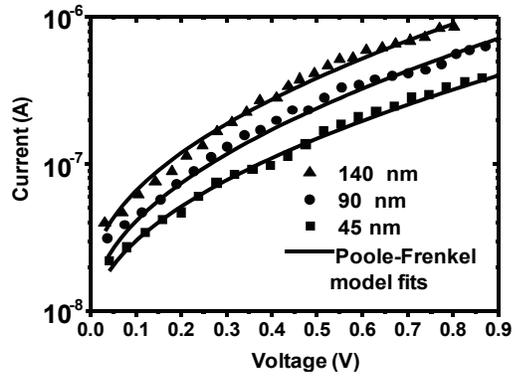

Figure 4, Mitra et al.